\documentclass[11pt,a4paper]{article}

\newcommand{\eq}{\begin{equation}}
\newcommand{\eeq}{\end{equation}}
\newcommand{\eqa}{\begin{eqnarray}}
\newcommand{\eeqa}{\end{eqnarray}}

\newcommand{\dee}{{\rm d}}
\newcommand{\infint}{\int_{-\infty}^{\infty}}
\newcommand{\sech}{{\rm sech}}
\newcommand{\cosech}{{\rm cosech}}
\newcommand{\infsum}{\sum_{n=-\infty}^{\infty}}

\newcommand{\mathl}{{\mathcal L}}

\usepackage{times}
\usepackage{graphicx}
\usepackage{amsfonts}
\usepackage{bm}
\usepackage[margin=0.8in,top=0.7in,bottom=0.7in]{geometry}

\begin{document}



\title{Structure and dynamics of crowdion defects in bcc metals}

\author{SP Fitzgerald\footnote{
S.P.Fitzgerald@leeds.ac.uk}
\\Department of Applied Mathematics\\
University of Leeds, Leeds, UK}



\maketitle

\begin{abstract}

Crowdion defects are produced in body centred cubic metals under irradiation. Their structure and diffusive dynamics play a governing role in microstructural evolution, and hence the mechanical properties of nuclear materials. In this paper we apply the analytical Frenkel-Kontorova model to crowdions and clusters thereof (prismatic dislocation loops) and show that the Peierls potential in which these defects diffuse is remarkably small (in the micro eV range as compared to the eV range for other defects). We also develop a coarse-grained statistical methodology for simulating these fast-diffusing objects in the context of object kinetic Monte Carlo, which is less vulnerable to the low barrier problem than na\"ive stochastic simulation.  

\end{abstract}

Crowdions \cite{kosevich2006} are the most stable configuration of self-interstitial atomic defect in the body-centred-cubic (bcc) transition metals 
V, Nb, Ta, Cr, Mo and W \cite{derlet2007}. They are produced in large quantities under irradiation, and agglomerate into the prismatic dislocation 
loops that characterize radiation damage. They are distinguished from other defect configurations by their effectively 
one-dimensional nature (along a close-packed crystal direction; $\langle 111\rangle$ in bcc metals), and exhibit numerous 
interesting properties, most notably their extremely low migration barriers (of order meV). In the next section, we review the Frenkel-Kontorova / sine-
Gordon model for $\langle 111\rangle$ crowdions, and then discuss its extension to the more realistic double-sine potential. Then we derive the {\it Peierls potential}, i.e. the effective potential within which the defect diffuses through the crystal. We then discuss crowdion clusters (aka prismatic interstitial-type dislocation loops), and the profound differences between the Peierls potential experienced by loop and that experienced by isolated 
crowdions. Finally we consider the 3D diffusion of crowdions, which is characterized by fast, virtually free diffusion along close-packed $\langle 111\rangle$ directions, separated by occasional stochastic changes to other $\langle 111\rangle$ directions. We show that the anisotropy of the 
diffusion can be neglected on timescales larger than the inverse direction-changing rate, and suggest an efficient simulation algorithm.

\section{Frenkel-Kontorova model}

From a theoretical point of view, their most attractive feature is the analytical tractability afforded by their one-dimensional nature. Below 
we introduce the Frenkel-Kontorova model \cite{braun2004}, a versatile one-dimensional model for the treatment of crowdions and also dislocation lines. 
The starting point is the Lagrangian
\eqa
\mathl & = & \infsum \left\{ \frac{m\dot z_n^2}{2} - \frac{\beta}{2}
\left(z_{n+1}-z_n-a\right)^2 - V(z_n)\right\}, \nonumber\\
& \to & \infint \left\{\frac{ m}{2}\left(\frac{\partial u}{\partial t}\right)^2
-\frac{\beta a^2}{2}\left(\frac{\partial u}{\partial z}\right)^2 - 
V\left(u(z,t)\right)\right\}
\dee z
\label{eqn:FK}, 
\eeqa where the sum runs over the close-packed string containing one additional atom, which have mass $m$, position $z_n$, and 
are connected by harmonic springs with constant $\beta$. The interaction with the surrounding ``perfect'' lattice is encoded in the 
periodic potential $V(z_n)$. Assuming the atomic displacement $u_n \equiv z_n - na$ varies slowly with the atomic index $n$, it can be 
described by a continuous function $u(z,t)$, with boundary conditions $u(-\infty)=a,u(\infty)=0$, corresponding to the single additional 
atom in the string. $a$ is the equilibrium spacing, and is given by $r_0\sqrt 3/2 $ for the $\langle 111\rangle$ direction in a bcc crystal 
with lattice constant $r_0$. The simplest choice for the lattice potential is $V_0\sin^2\left(\pi z/a\right)$, and if we seek a static solution 
to the Euler-Lagrange equation corresponding to Eq.(\ref{eqn:FK}), we find 
\eq
u(z;z_0) = \frac{2a}{\pi}\tan^{-1}{\rm e}^{-\mu(z-z_0)}, \label{eqn:profile}
\eeq where $\mu^2 = 2\pi^2 V_0/\beta a^4$. This displacement profile smoothly varies from 0 to $a$ as $z$ goes from $-\infty$ to 
$\infty$, with the variation taking place over a lengthscale $1/\mu$. Thus $\mu$ encodes the width of the crowdion, reflecting the 
relative strengths of the intra-string ($\beta$) and surrounding lattice ($V_0$) interactions. $z_0$ is the crowdion centre-of-mass coordinate, 
i.e. its position in the $\langle 111\rangle$ string.

In the continuum limit, the energy of a static crowdion can be calculated by inserting the displacement profile Eq.\ref{eqn:profile} into the 
(static) Hamiltonian \cite{kosevich2006}
\eq
E_0 =  \infint \left\{
\frac{\beta a^2}{2}\left(\frac{\partial u}{\partial z}\right)^2 + 
V\left(u(z,t)\right)\right\}\dee z =  \left( \frac{\beta a^4\mu^2}{2\pi^2} + V_0\right)\frac{2}{a\mu} = \frac{2a}{\pi}\sqrt{2V_0\beta}.\label{eqn:E0}
\eeq Note how the two terms in the energy, corresponding to the intra-string ($\beta$) and surrounding lattice ($V_0$) interactions, are equal at every point. 
Also, $E_0$ is independent of $z_0$, and so is independent of position. This is an artefact of the 
continuum limit we have taken, and discreteness can be approximately reintroduced by assuming the crowdion's profile remains fixed as it moves 
through the crystal, and exploiting the equipartition of the energy between string and lattice to write 
\eq
E_{\rm discrete} = \infsum \left( \frac{\beta}{2}\left(z_{n+1} - z_n - a\right)^2 + V(z_n)\right) \to 2\infsum V(u_n); \; u_n = u(na),  
\eeq {\it i.e.} the continuum solution is evaluated at each discrete atom. The Poisson summation formula then leads to a Fourier series for 
the {\it Peierls potential} for the defect: 
\eq
E = E_0 + \frac{2V_0\pi^2}{\mu^2 a^2}\sum_{n=1}^{\infty}\, n\, \cos\frac{2\pi n z_0}{a}\,\,
\cosech\frac{\pi^2 n}{\mu a}.
\eeq This is the potential in which the defect moves, and the $\cosech ({\pi^2 n}/{\mu a})$ factor strongly suppresses its magnitude 
when $\mu a < 1$, which is the case for crowdions. This is delocalization: the intra-string interaction is greater than the lattice interaction, 
meaning the displacement is spread over many atoms. Moving the defect centre-of-mass one lattice parameter corresponds to tiny motions 
of many atoms, leading to a suppressed migration barrier. The first term in the series is adequate, and 
\eq
E_{\rm mig}\approx\frac{8V_0\pi^2}{\mu^2 a^2}{\rm cosech} \frac{\pi^2}{\mu a},
\eeq which is in the $\mu$eV range for reasonable values of the parameters (see \cite{fitzgerald2008b} $V_0\sim$ 1eV, $\beta a^2\sim$ 50-100 eV).

\section{Double sine-Gordon model}

\begin{figure}
\centering
\includegraphics[width=0.45\textwidth]{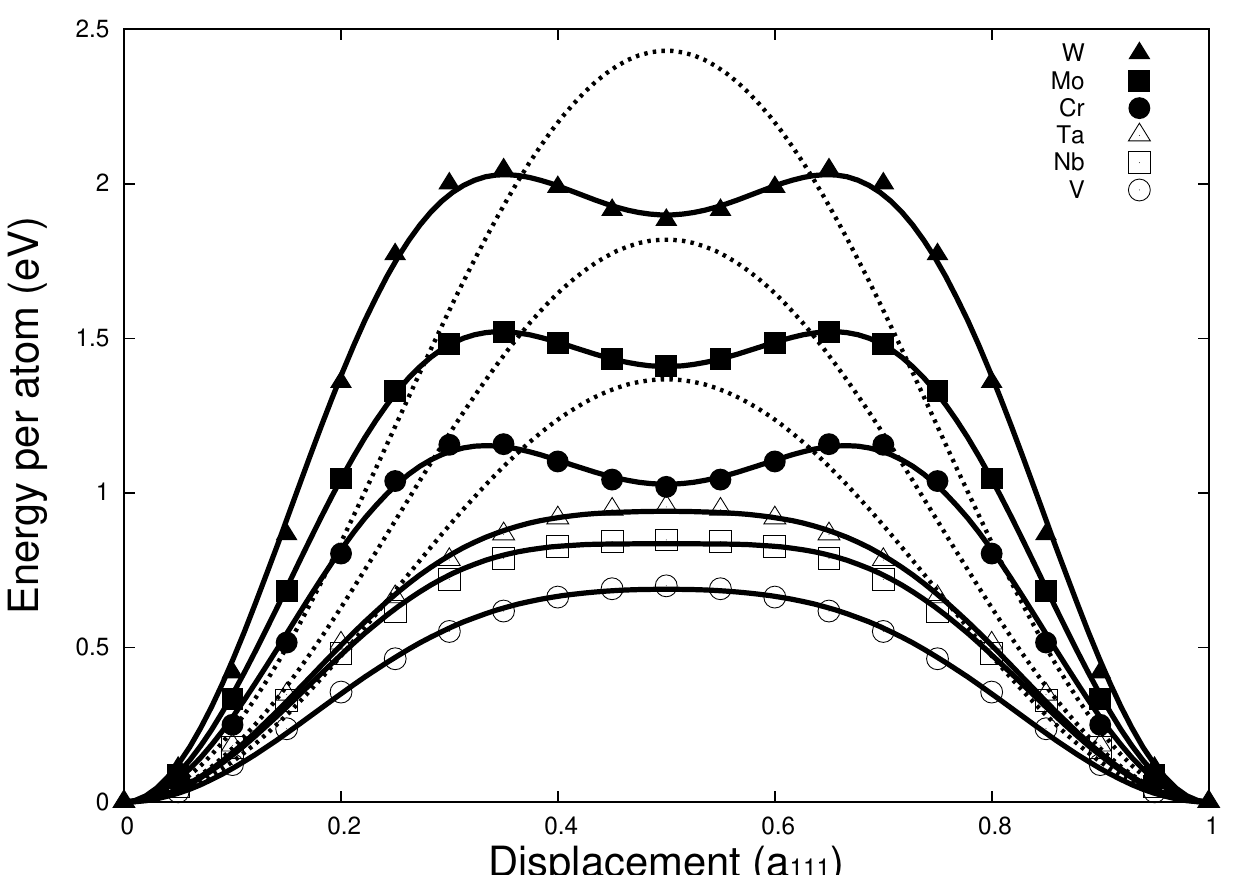}
\includegraphics[width=0.45\textwidth]{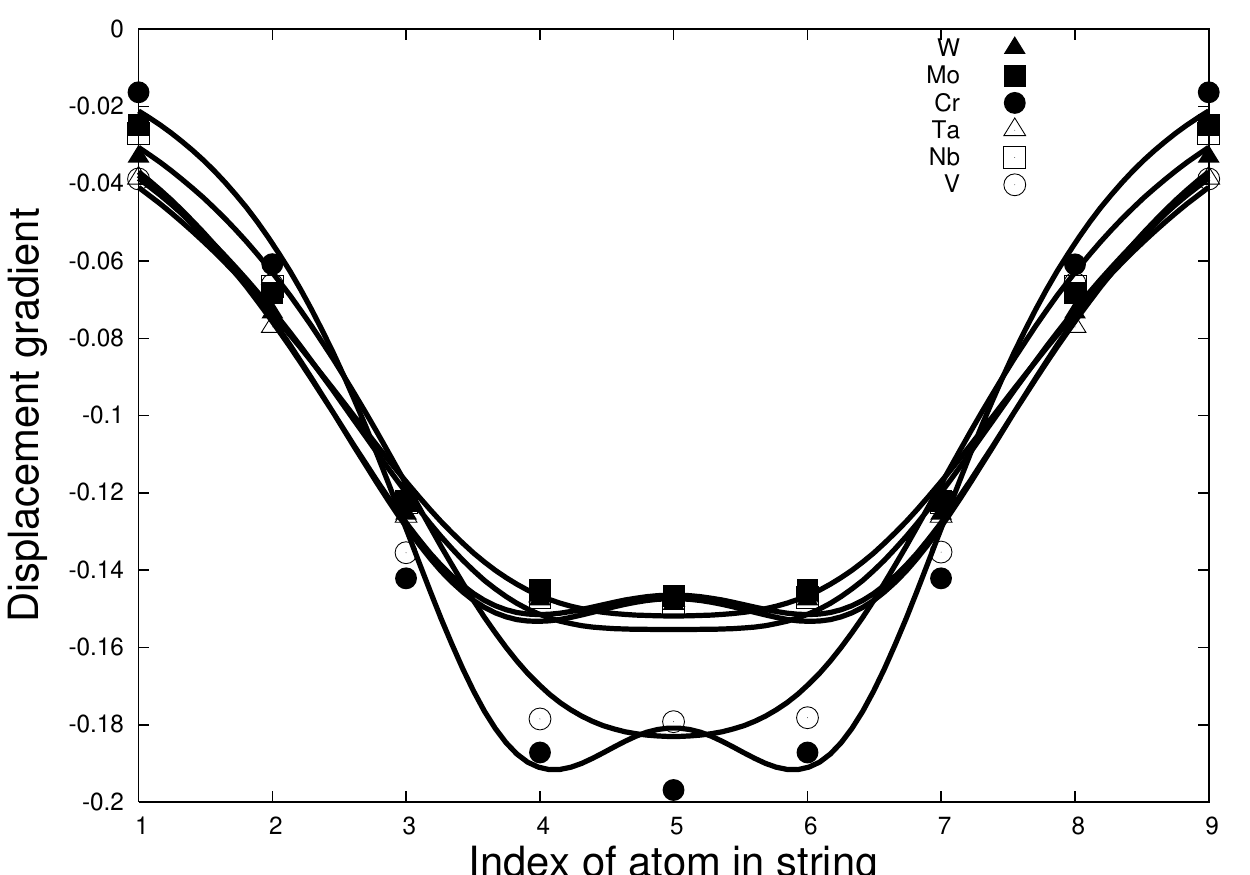}
\caption{Lattice potential (left) and atomic displacement gradients (right) for crowdions in the bcc transition metals (DFT; data from \cite{fitzgerald2008b}). Solid lines: double sine fits; dashed line: single sine fit.}
\label{fig:potdisp}
\end{figure}

Atomistic simulations \cite{derlet2007} suggest that, whilst very low, the crowdion migration barrier is in the meV rather than $\mu$eV range,
 indicating that 
the model described above is not the whole story. In fact, the assumption that the lattice potential is sinusoidal is not always accurate, as 
density functional calculations show. Particularly for the group VI metals Cr, Mo and W, the potential shows a local minimum midway between 
the main $a$-period minima, as can be seen in Fig. \ref{fig:potdisp}, \cite{fitzgerald2008b}.
\begin{figure}
\centering
\includegraphics[width=0.85\textwidth]{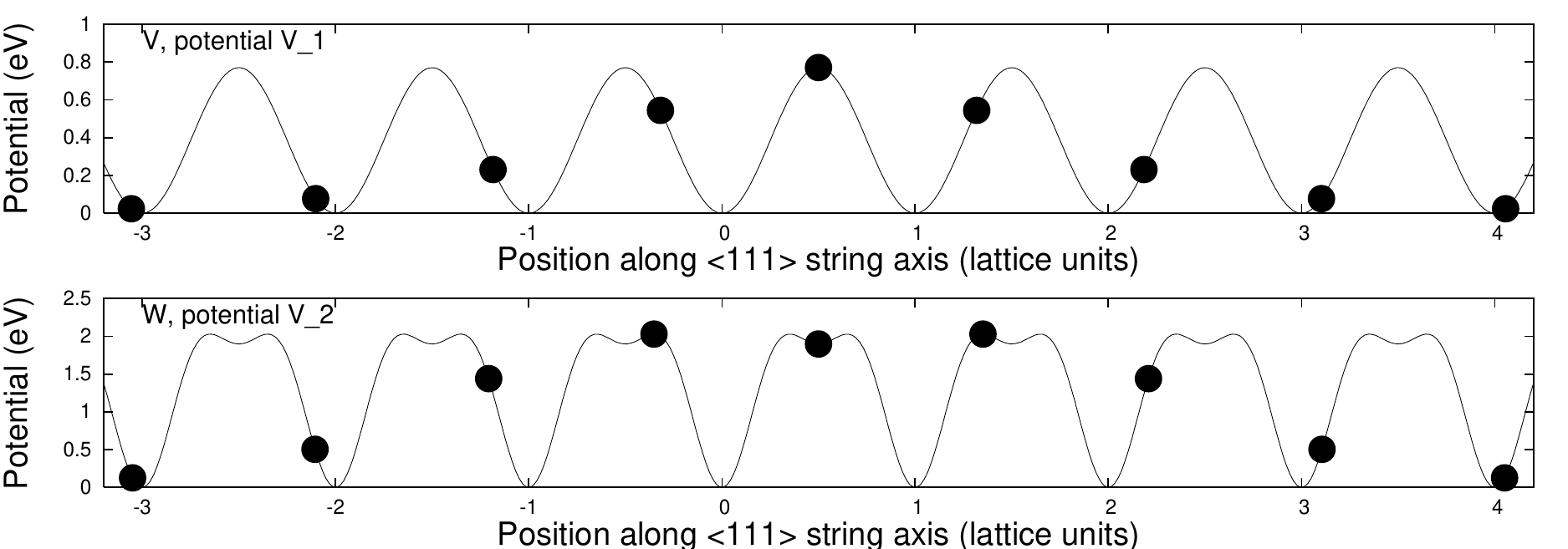}
\caption{Atomic positions for crowdions in the single- (top) and double-sine (bottom) models. Parameters are for vanadium and tungsten respectively.}
\label{fig:posns}
\end{figure}

These curves can be well-fitted by a double-sine potential 
\eq 
V(z)= V_0\left(\sin^2\left(\vphantom{\frac{2\pi z}{a}} \frac{\pi z}{a}
\vphantom{\frac{2\pi z}{a}} \right)
+ \frac{\alpha^2 - 1}{4}\sin^2\left( \frac{2\pi z}{a}\right)
\right),\label{eqn:V}
\eeq and the analysis carries forward, leading to a displacement solution
\eq
u(z;z_0) = \frac{a}{\pi}\arctan\left[ \frac{\alpha}{\sinh\left(\mu
\alpha(z-z_0)\right)}\right],
\label{eqn:disp}
\eeq and the width of the crowdion is now encoded by the combination $\mu\alpha$. A similar, yet more involved, calculation 
yield the Peierls potential 

\eq
E(z_0) = E_0 + \sum_{j=1}^{\infty}I_j\cos\left(\frac{2\pi j z_0}{a}\right),
\label{eqn:exp}
\eeq where

\eq
I_j  =  \frac{2V_0\alpha\pi}{\mu a}\cosech\left(\frac{\xi\pi}{2}\right)
\times \left\{\xi
\cos\left(\frac{\xi}{4}\ln\frac{q_+}{q_-}\right) \right.
 \left.  - \frac{1}{\alpha
\sqrt{\alpha^2-1}}\sin\left(\frac{\xi}{4}\ln\frac{q_+}{q_-}\right)\right\},
\label{eqn:I}
\eeq and $\xi = 2\pi j/\alpha\mu a$ and $q_{+,-} = 1-2\alpha^2 
\pm 2\alpha\sqrt{\alpha^2 - 1}$. The input parameters can be determined from density functional 
calculations, and the results for the migration 
barrier heights for V, Nb, Ta are 
$6.8\times 10^{-4}, 0.25\times 10^{-4}$ and $0.087\times 10^{-4}$ eV 
respectively, and those for Cr, Mo, W are $12\times 10^{-3}, 2.4\times 
10^{-3}$ and $2.6\times 10^{-3}$ eV respectively. A clear group-specific trend emerges, with the group VI 
metals having a deeper local minimum, and hence a larger migration barrier, than their group V counterparts. 
Still, all these barriers are remarkably low. 

%
%
%

%
\section{Multi-crowdion solutions}

\begin{figure}
\centering
\includegraphics[width=0.75\textwidth]{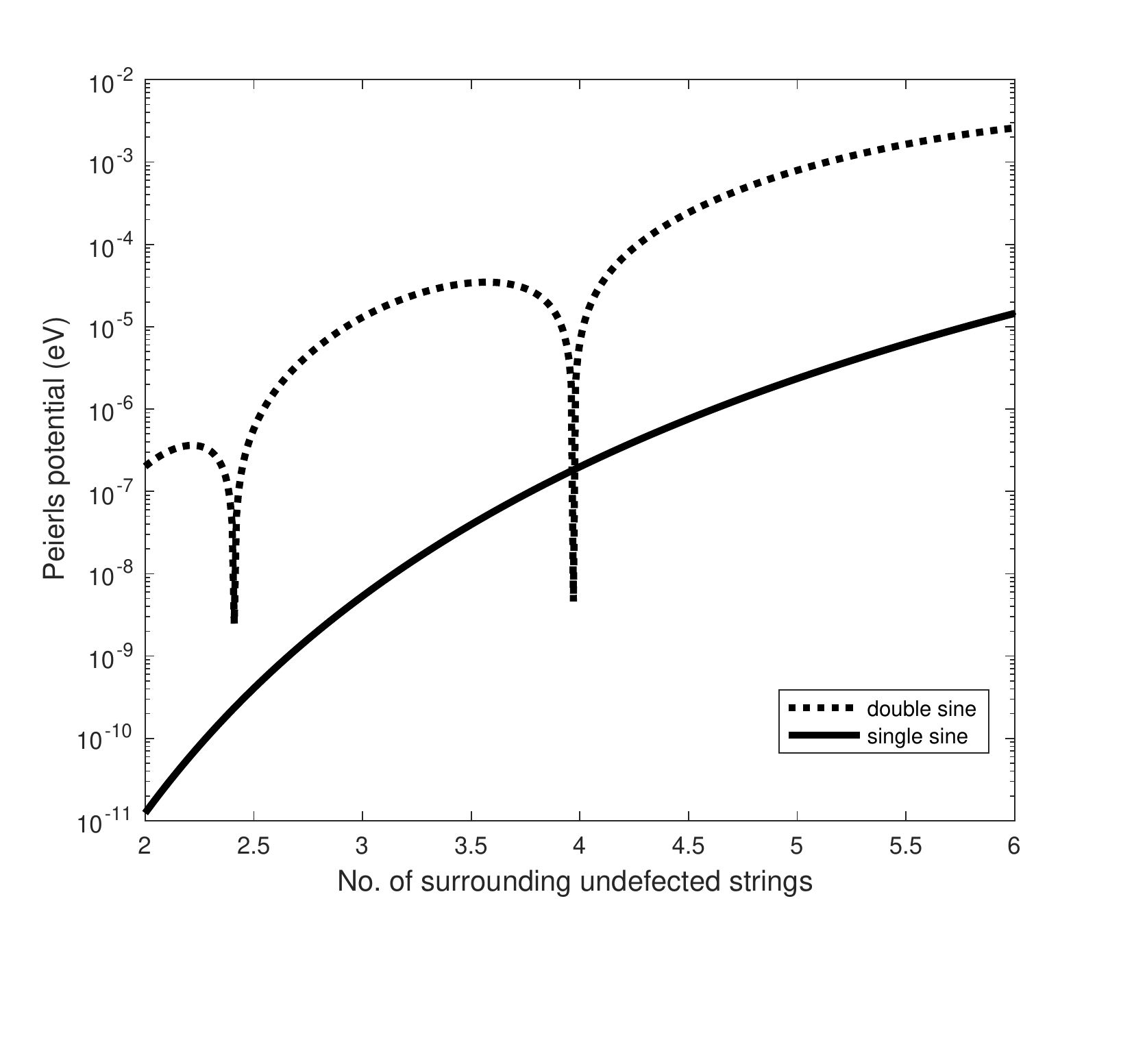}
\caption{Effect of number of undefected neighbour strings on crowdion Peierls potential. An isolated crowdion has 6, whereas most 
boundary crowdions in a cluster have 2.}
\label{fig:peierls}
\end{figure}

Crowdions cluster together to form $\bm{b} = \frac1{2}\langle 111\rangle$ prismatic dislocation loops, and the Frenkel-Kontorova 
model can be extended to treat these clusters \cite{dudarev2003}. Using the single-sine form for simplicity, the interaction potential between 
two crowdions in neighbouring 
parallel $\langle 111\rangle$ strings with displacement fields $u_0 = u(z;0),u_1=u(z;x)$ can be written 
\eqa
E_{\rm int}(x) & = & \infint \frac{V_0}{6}\sin^2\left(\frac{\pi}{a}\left(u_0 - u_1\right)\right){\rm d}z\nonumber \\
& = & \frac{2V_0}{3\mu}\tanh \frac{\mu x}{2}\left( \frac{\mu x}{2} \sech^2 \frac{\mu x}{2} + \tanh \frac{\mu x}{2}\right),
\eeqa where $x$ is the separation between the crowdions' centres of mass \cite{fitzgerald2015crowdion}. The factor of $1/6$ arises because $V_0$ was defined 
as the lattice potential for an isolated crowdion, surrounded by 6 neighbours. Each member of a crowdion pair has 5 undefected 
neighbour strings, so its $\mu \to \sqrt{5/6}\mu$ compared to an isolated crowdion. This small correction has important effects due to the 
extreme nonlinearity of the Peierls potential. For tungsten, the 2-crowdion interaction potential above is a slight (maximum 0.3eV) 
repulsion for large distances, and an 
attractive well when the separation is less than about 12 atomic spacings. The well depth is $\sim$3eV (DFT gives somewhat 
less than this \cite{marinica2013}, but the agreement for the single sine model is reasonable), so crowdions bind strongly together. The consequence for 
their displacement profile is that their $\mu$ is reduced, and hence they are more spread out down the $\langle 111\rangle$ 
string. For large clusters, only crowdions near the edge experience strong interactions with the undefected lattice. Crowdions in the 
interior are delocalized to such an extent that they are indistinguishable from perfect lattice, and the cluster becomes a prismatic 
dislocation loop, with strain localized to the perimeter. At the perimeter, each boundary crowdion has 2 or 3 undefected neighbour strings 
(depending on the geometry of the loop -- small $\bm{b} = \frac1{2}\langle 111\rangle$ loops are typically hexagonal, so ``corner'' crowdions 
have 3 perfect neighbours, whilst ``edge'' crowdions have 2). Fig.\ref{fig:peierls} shows the effect this has on the Peierls potential for crowdions in 
tungsten. The 
enhanced delocalization reduces the Peierls potential by at least 4 orders of magnitude, rendering it zero to all intents and purposes. This 
suppression comes again from the cosech$(.../\mu)$ term, which is an extremely nonlinear function of $\mu$. This 
completely outweighs the increased number of boundary crowdions experiencing the Peierls potential\footnote{Loops would need to contain 
several million defects to have the $>$10,000 boundary crowdions required.} therefore prismatic dislocation loops 
can move through the crystal effectively unimpeded, more easily even than isolated crowdions.

\section{3D diffusion of single crowdions}

Most defects migrate stochastically through the crystal with a diffusivity $D$ that takes the form $D = D_0\exp(E_{\rm mig}/k_{\rm B}T)$, 
corresponding to hops through the lattice that occur with an Arrhenius rate proportional to $\exp(E_{\rm mig}/k_{\rm B}T)$ ($T$ is the temperature, 
$E_{\rm mig}$ is the migration barrier and $k_{\rm B}$ is Boltzmann's constant). This expression depends on the implicit assumption that 
$E_{\rm mig}\gg k_{\rm B}T$, i.e. the hops are rare events. This clearly does not apply to crowdions for all but cryogenic temperatures. Indeed, 
for $E_{\rm mig}\ll k_{\rm B}T$, the diffusion is effectively free. For 1D motion in a sinusoidal potential, an exact solution for the hop rate exists 
for all temperatures, see e.g. \cite{swinburne2013}. 

Molecular dynamics simulations \cite{derlet2007} confirm the fast 1D nature of crowdion migration, but also show the defects changing from one $\langle 111\rangle$ 
direction to another. This occurs at a slower rate, comparable to the ``rare event'' hops of other crystal defects. This allows the crowdion to 
explore the entirety of the crystal, and in this section we calculate the effect on 3D diffusion, and outline a Monte Carlo algorithm for its simulation. 

Firstly assume that the direction-changing transition is a Poisson process with rate $\Gamma$. Then the time intervals between changes of direction 
will be exponentially distributed, with pdf $\psi(t) = \Gamma\exp(-\Gamma t)$. If we further assume that, during the time interval $t$ spent between 
direction changes, the crowdion diffuses normally with diffusivity $D$, then the hop lengths $x$, conditioned on the time interval $t$, will have the 
normal distribution $\Lambda(x|t) = \exp(-x^2/2Dt)/\sqrt{2\pi Dt}$. Since the hops are independent, we can reorder the series of hops, 
treat each $\langle 111\rangle$ direction independently in 1D, and project onto 3D space at the end. The fact that the directions along which the crowdion can diffuse are linearly dependent is immaterial, as shown below. 

In the bcc lattice, there are four (unsigned) $\langle 111\rangle$ directions along which crowdions can move, with unit vectors $\bm{\hat e_{1,2,3,4}}$. 
The final position of the crowdion is $\bm{x_f} = \sum_{i=1}^4s_i\bm{\hat e_i}$, where $s_i$ is the sum of the signed hop lengths in the $i$ direction. 
Each of these hop lengths is normally distributed with zero mean and variance $D\Delta\! t$ (and the $\Delta\! t$s are exponentially distributed, though 
that is not required). The total time $t=  \sum_{i=1}^4t_i$ where $t_i$ is the time spent hopping in each direction, i.e. the sum of the $\Delta\! t$s for each 
direction. The expected value for $|\bm{x_f}|^2$ is given by
\eqa
 {\mathbb E}\left(|\bm{x_f}|^2\right) =   {\mathbb E}\left(\sum_{i=1}^4s_i\bm{\hat e_i}\right)^2 & = &  |\bm{\hat e_i}|^2 {\mathbb E}\left(s_1^2\right) + ... 
 + 2\bm{\hat e_1\cdot\hat e_2} {\mathbb E} (s_1s_2) + ... \nonumber\\
 & = & Dt_1 + Dt_2 + Dt_3 + Dt_4 + 0\nonumber\\
 & = & Dt,
\eeqa where in the second line we used the fact that variances add in sums of normally distributed random variables, and that 
${\mathbb E}(s_1s_2) = {\mathbb E}(s_1){\mathbb E}(s_2) = 0$ by independence. The $\bm{\hat e_i}$s need not be orthogonal.

\begin{figure}
\centering
\includegraphics[width=0.24\textwidth]{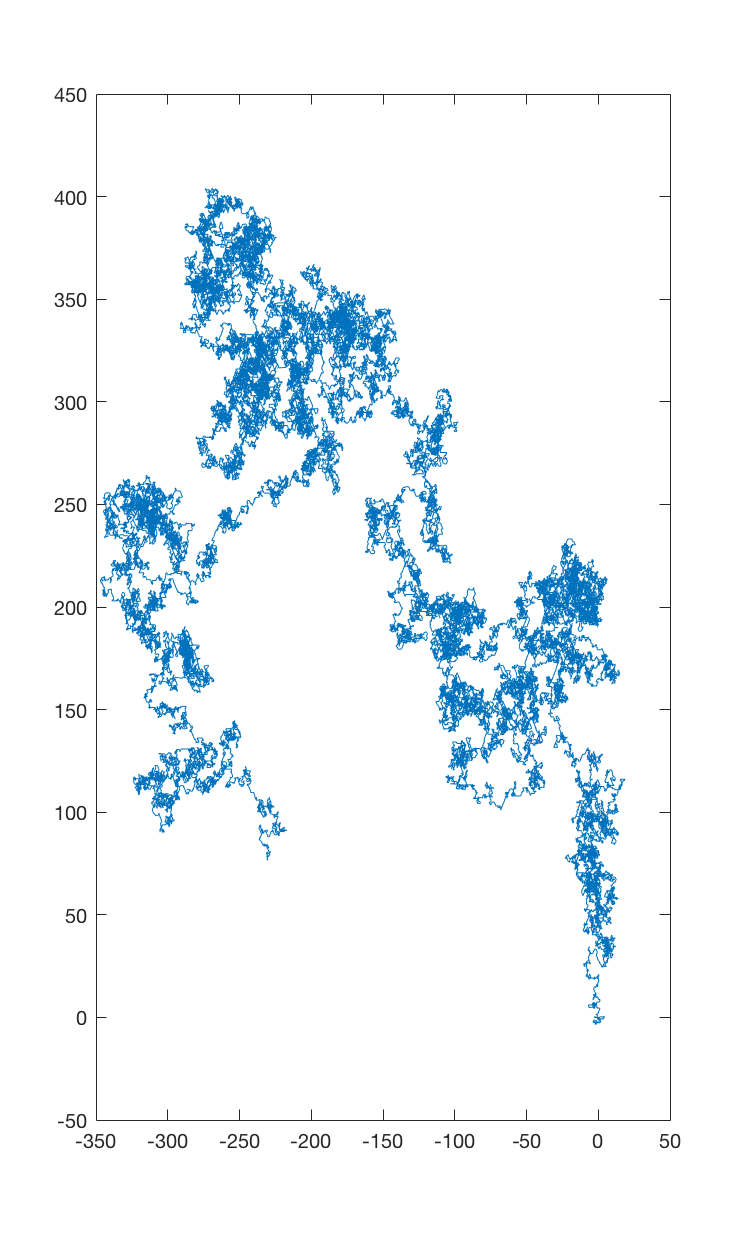}
\includegraphics[width=0.24\textwidth]{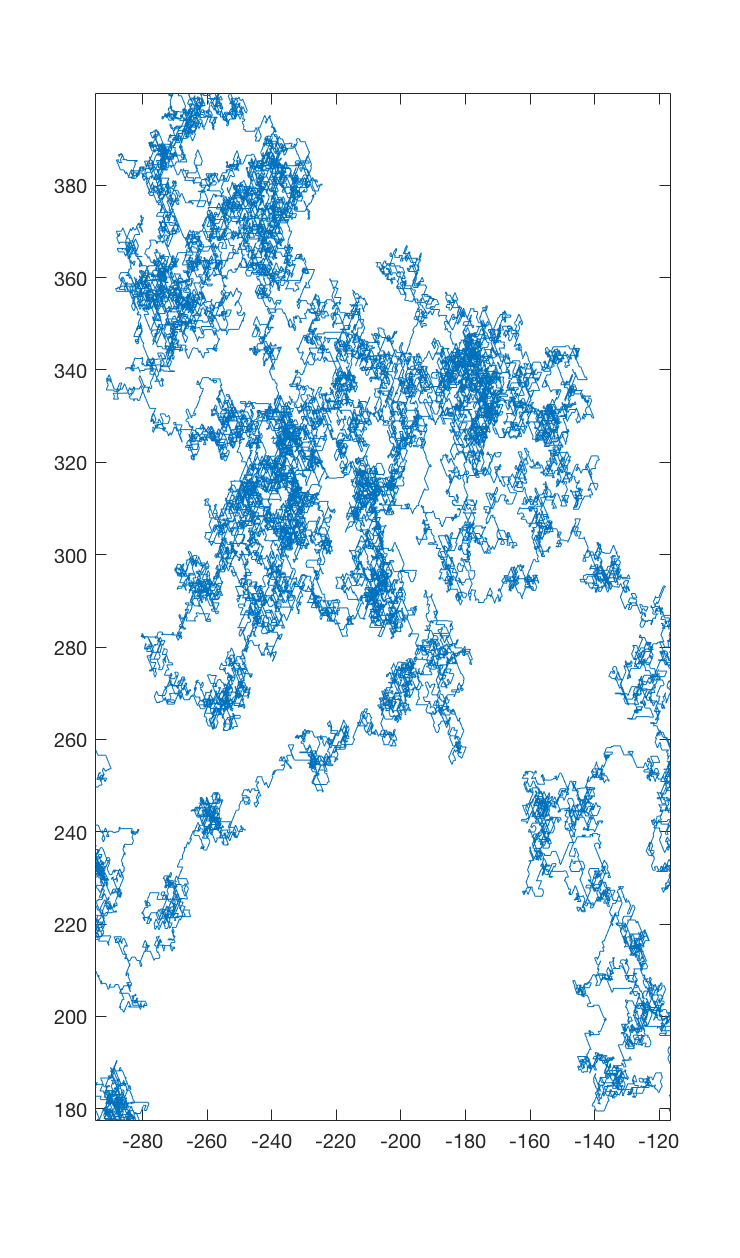}
\includegraphics[width=0.24\textwidth]{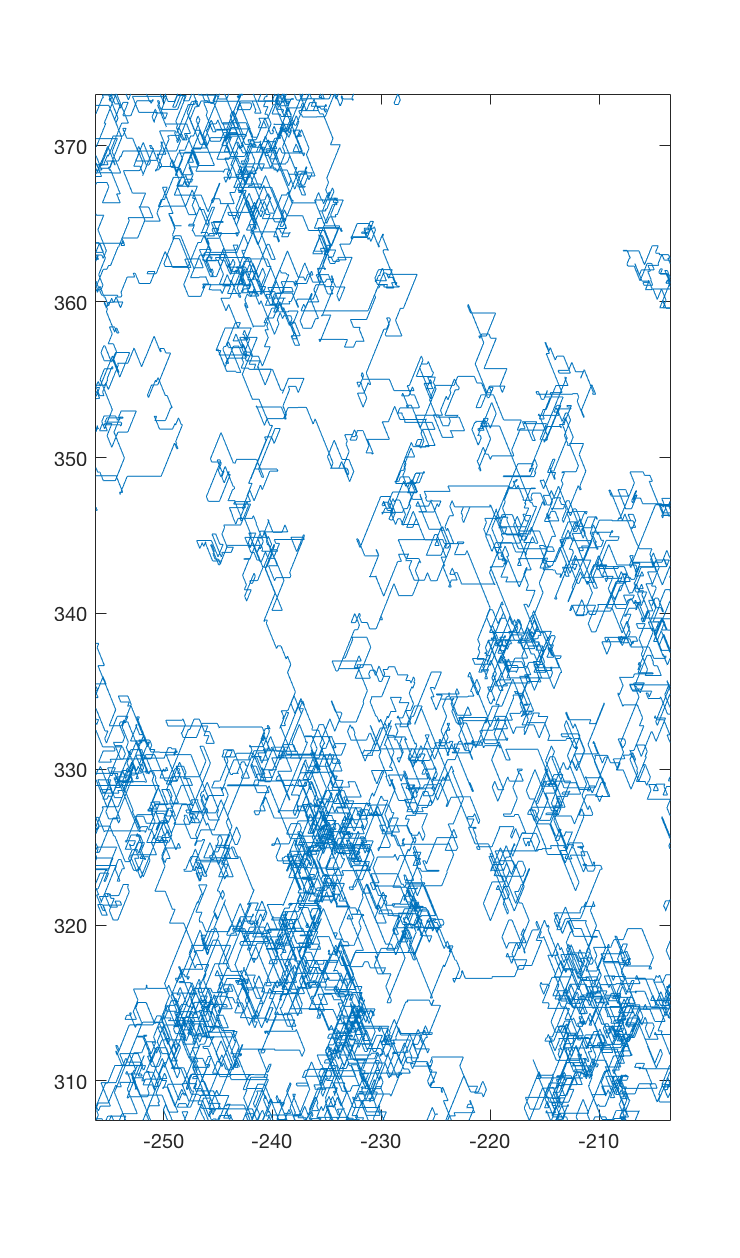}
\includegraphics[width=0.24\textwidth]{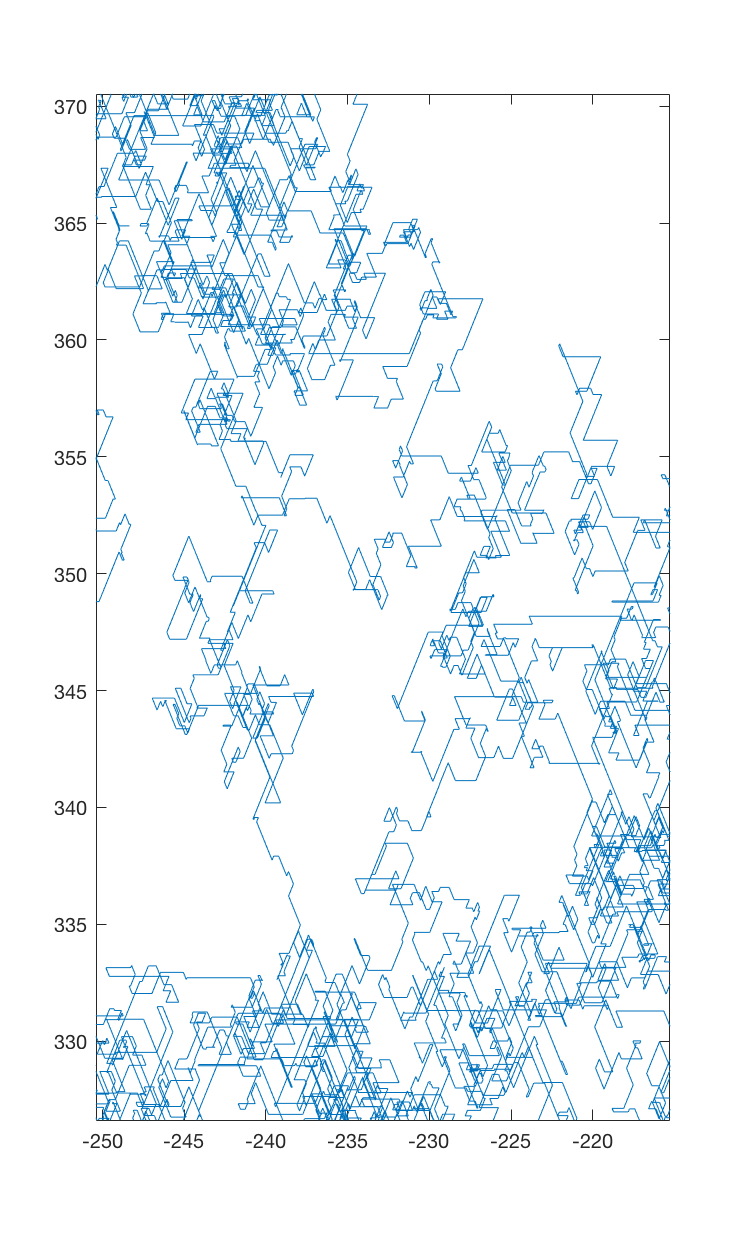}
\caption{Left to right: increasing magnification views of an example trajectory from crowdion Monte Carlo. Only at the smallest scales is the 
anisotropy of the diffusion evident. The time $t$ spent on a particular $\langle 111\rangle$ direction is drawn from an 
exponential distribution, then the distance diffused on that direction prior to the change is drawn from a normal distribution with variance 
$Dt$.}
\label{fig:sim}
\end{figure}

With the above assumptions, the pdf $W$ for the crowdion position $x$ at time $t$ satisfies the Chapman-Kolmogorov equation \cite{montroll}
\eq
W(x,t) = \int_0^t\int_{-\infty}^{\infty}\psi(t-t')\Lambda(x-x'|t-t')W(x',t')\dee x'\dee t' + \left(1-\int_0^t\psi(t)\dee t\right)W(x,0).
\eeq This reflects the sum over all possible hop lengths and times, and the second term is the probability density for the 
particle remaining at $x = 0$ until time $t$, $W(x,0) = \delta(x)$. Inserting the above forms for $\psi$ and $\Lambda$ then taking Fourier 
transforms in $x$ and Laplace transforms in $t$, $W(x,t)\to W(k,s)$, leads to 
\eq
W(k,s) = \frac{2s + \Gamma  k^2D}{(s+\Gamma)(2s+k^2D)}.
\eeq Now, since 
\eq
\frac{\partial^2 W(k,t)}{\partial k^2}\equiv \infint {\rm e}^{ikx}(-x^2) W(x,t)\dee x,
\eeq we can differentiate $W(k,s)$ twice with respect to $k$ and set $k=0$ to get (minus) the Laplace-transformed expected value for $x^2$. Inverting the 
transform gives
\eq
\langle x^2\rangle = D\left(t - \frac{1-\exp(-\Gamma t)}{\Gamma}\right)\sim Dt \;{\rm when}\; t\gg\frac1{\Gamma}.
\eeq So for sufficiently large times, the effective diffusivity is that of the 1D fast motion, but how long until this approximation is reasonable is controlled 
by the rate of direction changes, $\Gamma$. Indeed, for $t\ll 1/\Gamma$, $\langle x^2\rangle\sim D\Gamma t^2/2$. The MD simulations of \cite{derlet2007} give a rate 
\eq
\Gamma = 6.59\times 10^{12} \exp(-0.385 \,{\rm eV}/k_{\rm B}T)/{\rm sec}
\eeq for crowdions in tungsten, whereas the migration energy for vacancies is found to be 1.78eV. This suggests that, on the timescale of vacancy 
diffusion, crowdion diffusion is effectively isotropic, and the 1D nature of hops can be neglected. Crowdion clusters/prismatic loops, on the other hand, 
stick to single $\langle 111\rangle$ directions for much longer. Whilst rotations for very small loops are not impossible \cite{arakawa2006changes}, the activation energy is much higher. 

Stochastic computer simulations are most efficient when the events being sampled have rates as similar as possible. A kinetic Monte Carlo simulation 
of, say, crowdion and vacancy hopping would spend the vast majority of its time moving crowdions since their barriers are so low compared to those for 
vacancies (this is known generically as the low barrier problem). A more efficient approach would be to sample the direction-changing events, and then 
draw the crowdion's 1D motion from a normal distribution with appropriate time-dependent variance, as in the analytical approach above. Fig. 
\ref{fig:sim} shows an example trajectory from a million step simulation of this type, which can be performed in under a minute on an ordinary laptop. 
The 1D $\langle 111\rangle$ hops are only apparent when `zoomed in', and at larger scales are indistinguishable from standard diffusion. Indeed, 
given that crowdions' diffusion rate is typically many orders of magnitude higher than any other species', it may be advantageous to treat the crowdions 
using a density functional, in analogy with the DFT approach to electrons. 

\section{Conclusions}

In this paper, we have derived the surprising result that clusters of crowdions (aka prismatic dislocation loops) can move through a bcc crystal 
lattice virtually unimpeded (aside from dissipation). The periodic (Peierls) potential in which they move is fractions of a micro eV: several orders of 
magnitude lower than 
even that for an isolated crowdion. The reason for this is {\it delocalization} -- the lattice displacement induced by the additional atoms is spread 
over many atoms, meaning the translation of its centre of mass corresponds to the tiny motions of many more atoms. This is analogous to how the 
existence of dislocations allows the plastic deformation of crystals at far lower applied stresses than their ``theoretical strength'' would suggest.  

We then showed that the highly 
anisotropic diffusion of crowdions, which atomistic simulations have demonstrated, can be safely neglected at timescales sufficiently far above 
the timescale for direction changes. This will aid the development of hybrid mesoscale Monte Carlo simulations of defect structure evolution, by avoiding 
the low barrier problem associated with the suppression of the Peierls potential. 


\section*{Acknowledments}

SPF thanks Dr D Nguyen Manh and Dr M-C Marinica for many helpful discussions. This work was supported in part by the UK EPSRC, Grant 
number EP/R005974/1.

\end{document}